\documentclass[prl,showpacs,floatfix,amsmath,amssymb,superscriptaddress]{revtex4}
\pdfoutput=1
\usepackage{graphicx}
\usepackage{natbib}
\usepackage{wasysym}
\usepackage{latexsym}
\usepackage{amssymb,amsfonts}
\usepackage{amsmath}
\newcommand{\av}[1]{\left\langle{#1}\right\rangle}

\def\be{\begin{equation}}
\def\ee{\end{equation}}
\def\ba{\begin{eqnarray}}
\def\ea{\end{eqnarray}}

\def\ve{\varepsilon}
\def\a{\alpha}

\def\D{\Delta}

\def\t{\theta}

\begin{document}

\title{Mott physics and band topology in materials with strong spin-orbit interaction}

\date{\today}

\author{D. A. Pesin}
\email{pesin@physics.utexas.edu}
\altaffiliation[\newline Present address~: ]{Department of Physics, The University of Texas at Austin, Austin, TX 78712, USA}
\affiliation{Department of Physics, University of Washington,
Seattle, WA 98195, USA}
\affiliation{Kavli Institute for Theoretical Physics, Santa Barbara, CA 93106}
\author{Leon Balents}
\affiliation{Kavli Institute for Theoretical Physics, Santa Barbara, CA 93106}
\begin{abstract}
  Recent theory and experiment have revealed that strong
  spin-orbit coupling can have dramatic qualitative effects on the band
  structure of weakly interacting solids.  Indeed, it leads to a distinct
  phase of matter, the topological band insulator.  In this paper, we
  consider the combined effects of spin-orbit coupling and strong
  electron correlation, and show that the former has both
  quantitative and qualitative effects upon the correlation-driven Mott
  transition.  As a specific example we take Ir-based
  pyrochlores, where the subsystem of Ir $5d$ electrons is known to
  undergo a Mott transition.  At weak electron-electron interaction, we predict that Ir
  electrons are in a metallic phase at weak spin-orbit interaction, and
  in a topological band insulator phase at strong spin-orbit
  interaction. Very generally, we show that with increasing strength of the
  electron-electron interaction, the effective spin-orbit coupling is
  enhanced, increasing the domain of the topological band insulator.
  Furthermore, in our model, we argue that with increasing interactions,
  the topological band insulator is transformed into a
  ``topological Mott insulator'' phase, which is characterized by
  gapless surface spin-only excitations.  The full phase diagram also
  includes a  narrow region of gapless
  Mott insulator with a spinon Fermi surface, and a magnetically ordered
  state at still larger electron-electron interaction.
\end{abstract}  \maketitle

{\em Introduction:} The spin-orbit interaction (SOI), though apparently
a ``weak'' relativistic correction to the Schr\"odinger equation
(outside of high energy physics), is coming increasingly to the fore in
modern condensed matter physics. The discovery of Topological Band
Insulators (TBIs) in
theory~\cite{KaneMeleA,citeulike:5096324,BalentsMoore,FuKaneMele3D} and
experiment~\cite{Molenkamp,ZahidHasan} has revealed a surprising omission
in the ``textbook'' Bloch theory of the electronic structure of weakly
correlated solids. In these remarkable materials, strong spin-orbit
interactions allow a non-trivial topology of the electron bands,
resulting in protected ``helical'' edge and surface states in two and
three dimensional systems. Many other interesting phenomena, including
quantum number fractionalization and magneto-electric effects have been
predicted to occur in these systems, and are the subjects of a growing
experimental effort. In parallel, strong SOIs have been identified in a
growing variety of {\sl Mott} insulators, in which the insulating
behavior is driven by electron correlation rather than band structure.
For instance, SOIs are likely responsible for the large Wilson ratios
observed in many frustrated magnets at low
temperature~\cite{GangLB,tovar:024405}, and may be the driving force for
the formation of a ``spin-orbital liquid'' in some
Fe-spinels~\cite{GangSchnyderLB}. They have been experimentally shown to
control the orbital state in the Ir oxide Sr$_2$IrO$_4$ using resonant
x-ray scattering~\cite{citeulike:4746735}. A natural question is how
these two classes of phenomena are connected -- how does a material
progress from weak to strong correlation with strong SOIs? This is the
subject of the Mott transition with strong SOIs.

In the search for strong SOIs, one is driven to consider materials based
on heavy atoms (as SOI scales with atomic number as $Z^4$).  So far, the
experimental examples of topological insulators have been based on atoms
(Hg and Bi) in which the active electrons closest to the Fermi energy
inhabit s and p orbitals, which are generally weakly correlated.
However, strong spin-orbit coupling {\sl and} electron correlation may
be expected in compounds involving d electrons. Especially in 5d
transition metals, correlation effects are often of weak or intermediate
strength, making them perhaps competitive with those of SOIs and band
structure.  Due to the surge of activity in Ir-based transition metal
oxides~\cite{citeulike:4746735,shitade:256403,GangLB,matsuhira2007metal,okamoto2007spin,fukazawa2002filling,nakatsuji2006metallic}, we choose one of these as our model system, and study a simple
phenomenological Hamiltonian for the Mott transition in Ir pyrochlores.
We will see that this crystal structure naturally fosters TBI physics in
the weak-correlation limit, and therefore allows for a study of the most
interesting regimes of the Mott transition with strong SOIs.

\begin{figure}
\begin{center}
\includegraphics[width=0.5\columnwidth]{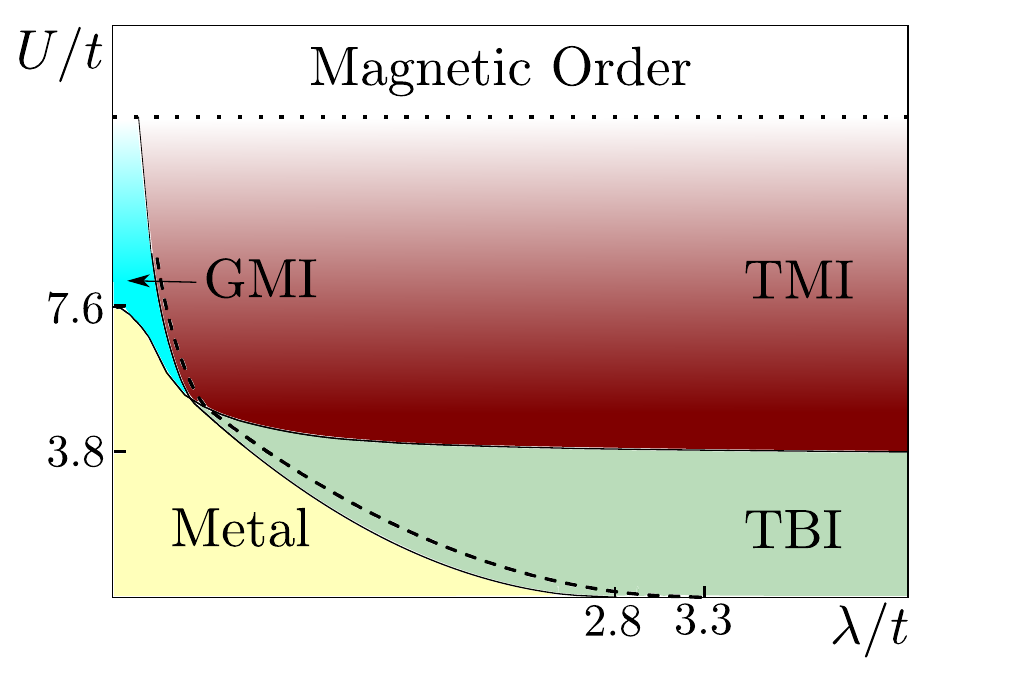}
\caption{Phase diagram based on the slave rotor approximation and strong
  coupling limit, as a function of Hubbard repulsion $U$ and spin-orbit
  coupling $\lambda$ (relative to hopping $t$).  The four main phases
  occuring for moderately strong electron-electron repulsion are a
  Metallic phase, Topological Band Insulator (TBI), Topological Mott
  Insulator phase (TMI), and Gapless Mott Insulator (GMI). The dashed
  line denotes an additional zero-gap semiconductor state due to an
  ``accidental'' gap closing. The dotted line schematically separates the
  large-$U$ region, where magnetic ordering is expected.  As discussed
  in the main text,  long-range Coulomb interactions are expected to
  induce an excitonic region in the vicinity of the  Metal-TBI boundary
  shown here.}
\label{fig:phased}
\end{center}
\end{figure}

Our main results are as follows.  Most importantly, we find that Mott correlations enhance the effects of SOIs, due to the suppression of effective bandwidth.  This leads to increased stability of the topological insulator state with interactions.  In addition, we suggest the existence of an intermediate {\em Topological Mott Insulator} (TMI) state, between the TBI and the familiar Mott insulating phases at stronger interactions.  The TMI phase may be regarded as a quantum spin liquid state, in which the electrons are effectively spin-charge separated, and neutral fermionic spin-$1/2$ spinons are deconfined.  The non-trivial band topology of the TBI is transferred to the spinons, which are thereby gapped in the bulk but gapless at the surface.  These results are summarized in the phase diagram in Fig.~\ref{fig:phased}.

\begin{figure}[hbtp]
\includegraphics[width=0.7\columnwidth]{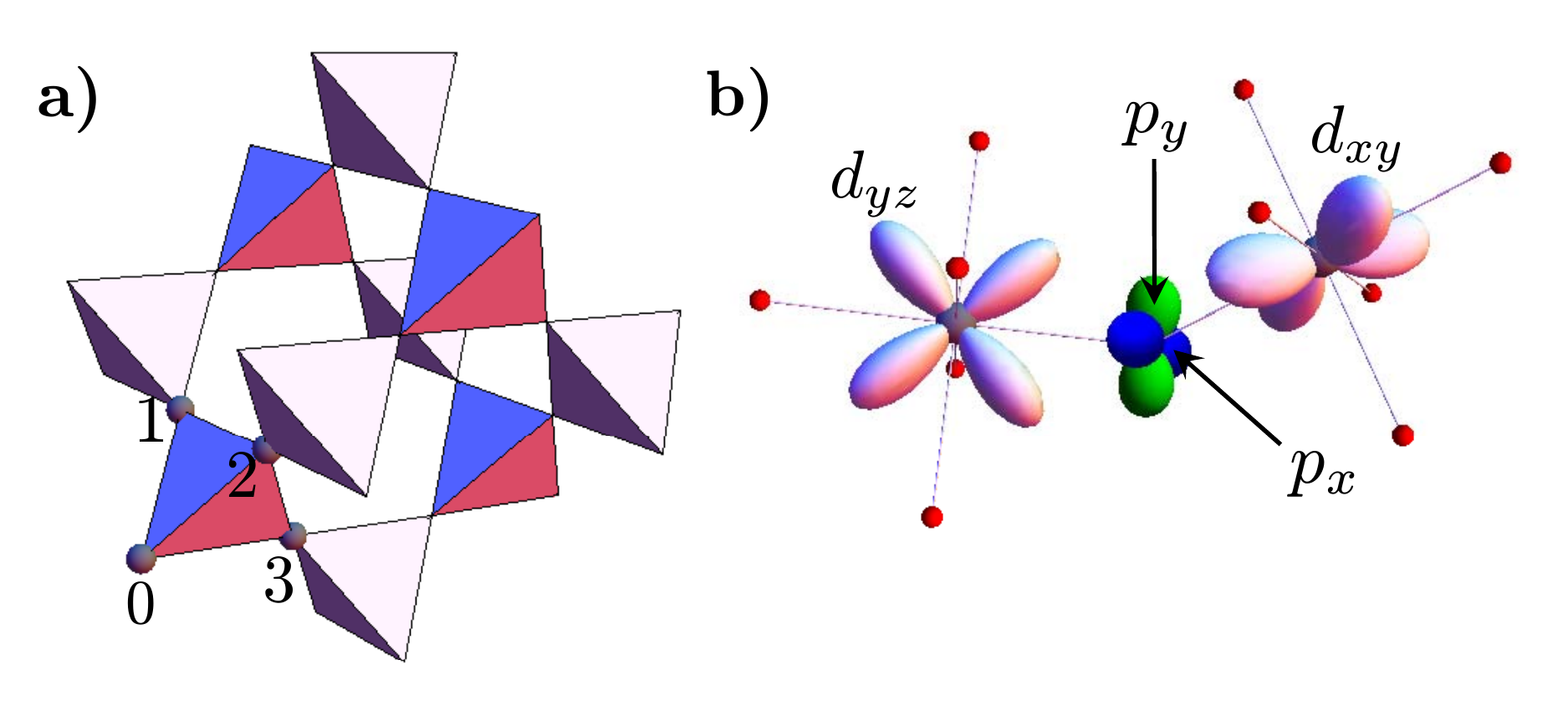}
\caption{{\bf a)}Pyrochlore lattice of corner-sharing tetrahedra. It can be viewed as the FCC lattice with tetrahedral bases added at each site. One of such bases, with Ir ions numbered from 0 to 3, is shown. {\bf b)} Oxygen mediated hopping between Ir sites. Sites 0 (on the left) and 3 (on the right) of the tetrahedral basis are shown (large gray spheres), together with their oxygen octahedral environment (small red spheres). On the ``shared'' oxygen site we show its $p_y$  orbital (green) with respect to the coordinate system of site 0, and $p_x$ orbital (blue) with respect to the coordinate system of site 3. Belonging to different coordinate systems, these orbitals are not orthogonal, the angle between them being $\approx 84^\circ$. Electrons can hop from local $d_{yz}$ orbital on site 0 onto $p_y$, and from the $d_{xy}$ orbital on site 3 onto $p_x$. Since the two p orbitals are not orthogonal, an effective Ir-Ir hopping is induced.}
\label{fig:IrEnv}
\end{figure}

{\em Model}: We begin by constructing a minimal microscopic model for the Mott transition in Ir-based pyrochlore oxides, A$_2$Ir$_2$O$_7$, where A is assumed to be an inert ion in the $3+$ state.  Therefore we consider $\textrm{Ir}^{4+}$ ions on the pyrochlore lattice, which is viewed as the FCC lattice with a tetrahedral basis, Fig.~\ref{fig:IrEnv}(a).  Each Ir ion is surrounded by an oxygen octahedron, Fig.~\ref{fig:IrEnv}(b).  In general, these octahedra are somewhat distorted, dependent upon the difference of the oxygen displacement parameter $x$ from the ideal value of $5/16$.  We neglect this small distortion (of 5 and 8 percent for A=Pr,Eu) here, in which case the Ir electrons experience approximately cubic $O_h$ symmetry, and the crystal field splits the five d-orbital states into (upper) $e_g$ doublet and (lower) $t_{2g}$ triplet. The value of the cubic crystal field splitting is assumed large compared to the Hund exchange energy, thus the latter is neglected. The on-site spin-orbit interaction thus is taken to act within the $t_{2g}$ manifold.  In local axes bound to the oxygen octahedron the $t_{2g}$ orbitals are $d_{yz}, d_{zx}$ and $d_{xy}$. In Supplementary Information we present the rotation matrices that turn the cubic axes into the local octahedral ones at each Ir site. The axes themselves are not shown in Fig.~\ref{fig:IrEnv} for clarity reasons.

Now consider the SOI, denoting its strength by $\lambda$.  It is well-known that, projected within the $t_{2g}$ manifold, the orbital angular momentum operator is equal to  {\sl minus} the usual spin-one angular momentum operator ${\boldsymbol {\sf L}}$~\cite{Abragam}.   Thus we may effectively consider the pseudo-total angular momentum ${\bf J} = {\boldsymbol {\sf L}}+{\bf S}$, which commutes with the SOI.  The multiplet structure for the usual $\ell=1$ states is inverted, and the $t_{2g}$ manifold splits into a $j=1/2$ doublet with energy $\lambda$, and a $j=3/2$ quadruplet with energy $-\lambda/2$.  We will work in the local diagonal basis of the $j$ eigenstates, and introduce a single label $\a$ such that $\a=1,2$ and $\a=3\ldots6$ denote the doublet and quadruplet, respectively, with orbital energy $\ve_\a =\lambda$ for the doublet and $\ve_\a=-\lambda/2$ for the quadruplet. The specific form of the wave functions of these states are given in the Supplementary Information.

We assume that hopping between nearest neighbor Ir ions is accomplished
via the oxygen atoms nearest to a given pair, Fig.~\ref{fig:IrEnv}(b). In reality, this is not
necessarily the case~\cite{JYucondmat}. However, our assumption
minimizes the number of free parameters, and is resilient to
perturbations that are not too strong. The model with oxygen-mediated hopping is preferable as it contains a single parameter determining the hopping strength: the hopping integral between Ir $t_{2g}$ states and O p-orbitals ($V_{pd\pi}$ in the terminology of Ref.~\cite{SlaterKoster}).  Integrating out the oxygens, and taking the simplest on-site Coulomb interaction involving the total charge only, we arrive at the Hubbard Hamiltonian,
\begin{eqnarray}\label{eq: Hamiltonian}
H&=&\sum_{Ri\a}(\ve_\a-\mu)d^\dagger_{Ri\a}d^{\vphantom\dagger}_{Ri\a}+t\sum_{\substack{\av{Ri,R'i'}\\ \a\a'}} T^{ii'}_{\a\a'}d^\dagger_{Ri\a}d^{\vphantom\dagger}_{R'i'\a'}
  +\frac{U}{2}\sum_{Ri}\left(\sum_{\a}d^\dagger_{Ri\a}d^{\vphantom\dagger}_{Ri\a}-n_d\right)^2,
\end{eqnarray}
where $R$ and $i$ label the sites of the Bravais lattice and the tetrahedral basis, and $n_d=5$ is the number of 5d-electrons on each $\textrm{Ir}^{4+}$ ion. The strength of the hopping is parameterized by a single energy scale $t \propto V^2_{pd\pi}/\D$, where $\D$ is the energy difference between the Ir d and O p states. The procedure to obtain the dimensionless hopping matrices $T^{ii'}_{\a\a'}$, arising from the Ir-O-Ir hopping path, taking into account the rotation between the local cubic axes of each Ir ion, is given in the Supplementary Information.

\begin{figure}
\begin{center}
\includegraphics[width=0.5\columnwidth]{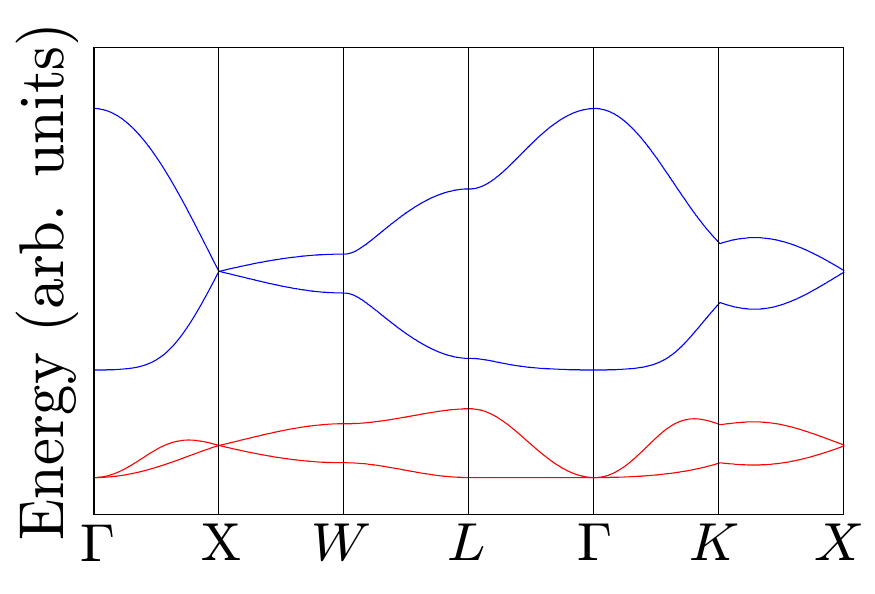}
\caption{Electronic band structure of Ir $5d$ electrons on the pyrochlore lattice at large spin-orbit coupling, $\lambda/t\to\infty$. Only the relevant four doubly degenerate bands are shown. A band gap between the filled lower two bands and empty upper two bands is clearly seen.}
\label{fig:bands}
\end{center}
\end{figure}

{\sl Band structure:} The Hamiltonian~(\ref{eq: Hamiltonian}) contains two dimensionless parameters: $\lambda/t$ and $U/t$, which define the phase diagram in Fig.~\ref{fig:phased}.  It is instructive to consider first various simple limits.
For $U=0$, we have a free electron model, which is of course exactly soluble.
Due to inversion symmetry, one obtains in general 12 doubly degenerate bands.  For small $\lambda/t$, these overlap at the Fermi energy and one obtains a metal.  For large $\lambda/t$, the upper 4 bands originating from the $j=1/2$ doublet become well-separated from the lower 8 bands.  Because there are four holes per unit cell, the upper 4 bands are half-filled in total.  On inspection, we see (Fig.~\ref{fig:bands}) that they exhibit a band gap, indicating the formation of a band insulator at large $\lambda$.

\begin{figure}
\begin{center}
\includegraphics[width=0.5\columnwidth]{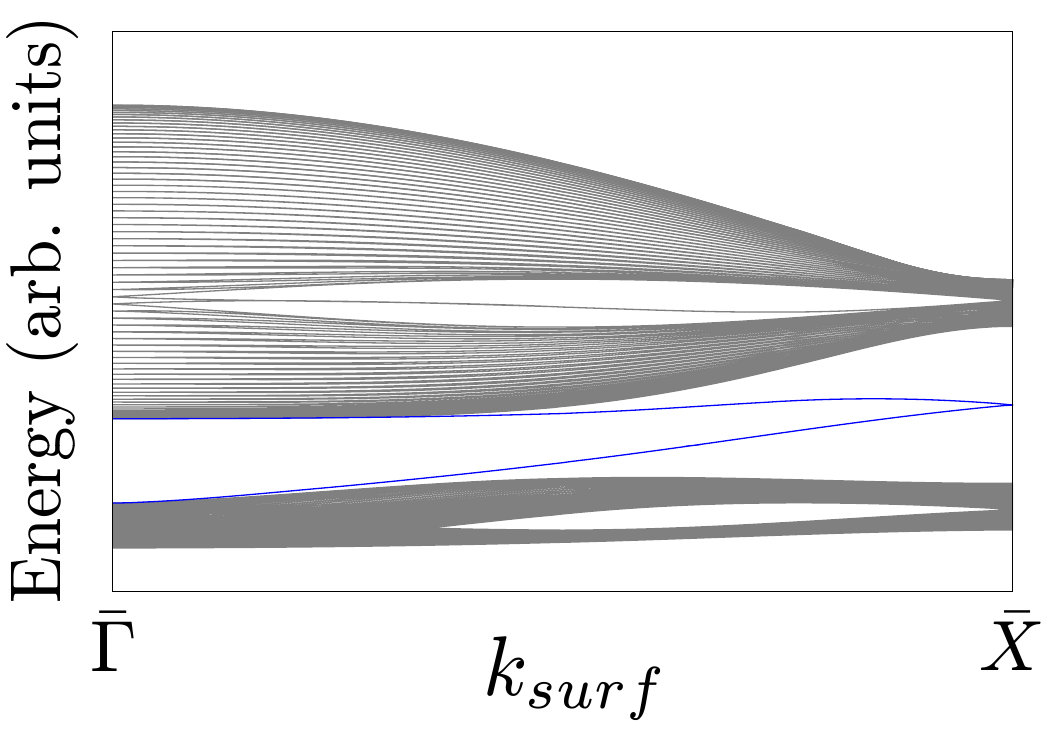}
\caption{Surface state spectrum. Projected bulk spectrum (gray), and the surface spectrum (solid blue line) for a $(010)$ surface are shown. The spectrum is plotted as a function of the surface wave vector $k_{surf}\equiv(k_x,k_z)$ running from $\bar{\Gamma}=(0,0)$ to $\bar{X}=(0,2\pi)$.}
\label{fig:surface}
\end{center}
\end{figure}

As shown by Fu and Kane~\cite{FuKaneTBIinv2007}, one can determine the band topology of an insulator with inversion symmetry either from the parity of the Hamiltonian eigenstates at time-reversal invariant momenta, or from the number of Dirac points on the surface of the insulator.   Applying the first criterion (see Supplementary Information), we find that the large $\lambda/t$ state is a pure ``strong'' TBI of the spinons (the weak $Z_2$ invariants vanish, consistent with cubic symmetry).   We also calculated the surface state spectrum (Fig.~\ref{fig:surface}), which shows the required odd number of intersections with the Fermi level on passing between time reversal invariant surface momenta.
The behavior for general $\lambda/t$ is as follows.  For $\lambda/t<2.8$, one obtains a metallic state, while for $\lambda/t>2.8$ the bands separate at the Fermi energy.  For almost all of this range of large $\lambda/t$, the system is a (strong) TBI.  However, an ``accidental'' closing of the band gap occurs at $\lambda/t \approx 3.3$, at which point it is a zero-gap semiconductor with 8 Dirac points located along the $\langle 111\rangle$ directions in reciprocal space.  Because the number of these Dirac points is even, there is no change in band topology due to the gap closure.

{\em Strong coupling limit:}  Now consider large $U/t$.  In this limit, one has a Mott insulator, and the Hamiltonian is effectively projected into the space of one hole per Ir site, and the system is described by a spin-orbit (Kugel-Khomskii type) model.  Superexchange leads to spin (and orbital) exchange of order $J \sim t^2/U \ll t$.  We see that in this limit SOI is only weak if $\lambda \ll J \ll t$.  Thus the strong SOI regime is greatly enhanced with increasing correlations, as the relevant ``bandwidth'' for large $U$ is exchange rather than hopping.  Since the general spin-orbital  Hamiltonian is cumbersome, and only relevant for very weak SOI, we will focus only on the strong SOI regime.  Here, only the half-filled doublet at each site is relevant, and the effective Hamiltonian is of Heisenberg spin-exchange type, with an effective spin $1/2$ at each site. The derivation of the low-energy Hamiltonian by second order degenerate perturbation theory is standard. It is customary to write the resulting Hamiltonian as a sum of isotropic exchange, Dzyaloshinskii-Moriya (DM), and anisotropic exchange parts (suppressing the  Bravais lattice index $R$ for brevity):
\begin{equation}\label{eq:spinHamiltonian}
H_{spin}=\frac{4t^2}{U}\sum_{ii'}\left[-\frac{19}{243}+J\vec{S}_i\vec{S}_{i'}
  +{\vec D}_{ii'}\cdot\vec{S}_i\times\vec{S}_{i'}
  +\vec{S}_i\cdot\overleftrightarrow{\Gamma}_{ii'}\cdot\vec{S}_{i'}\right]
\end{equation}
The convention to make the choice of $D_{ii'}$ unique is that $i'>i$. Since all $D$'s and $\Gamma$'s are related by symmetry, it is enough to specify them for one bond only:
\begin{eqnarray}\label{eq:Hspinprmtrs}
  J&=&\frac{49132}{177147},\nonumber\\
  D_{01}&=&\frac{7280}{59049}\left(0,1,-1\right),\nonumber\\
  \Gamma_{01}&=&\frac{1568}{59049}\left(
\begin{array}{ccc}
 -\frac{2}{3} & 0 & 0 \\
 0 & \frac{1}{3} & -1 \\
 0 & -1 & \frac{1}{3}
\end{array}
\right).
\end{eqnarray}
We note that the anisotropic exchange term, $\overleftrightarrow{\Gamma}_{01}$, is small compared to the other ones.  Spin ordering in this model was considered in Ref.~\cite{Elhajal}. Our DM term corresponds to the ``indirect'' case of Ref.~\cite{Elhajal}, and $|D_{01}|/J \approx 0.63$, i.e. very large DM interactions.  In this case, a magnetically ordered ground state is expected, which breaks point group symmetries but does not enlarge the unit cell.

{\em Slave-Rotor approach:} An exact or accurate numerical solution for the full phase diagram for Eq.~(\ref{eq: Hamiltonian}) is very challenging, especially at intermediate $U/t$.  To study it, we employ the slave-rotor approximation of Ref.~\cite{FlorensGeorges}.  This approach has a number of merits.  It becomes exact for $U/t=0$, and captures the bandwidth reduction with increasing $U/t$.  Its predictions for Hubbard models on other frustrated lattices without SOI at intermediate $U/t$ are in agreement with more controlled approaches such as the path integral renormalization group~\cite{mizusaki:014421}, Gutzwiller-type variational wavefunctions~\cite{motrunich2005vst}, and the variational cluster method~\cite{sahebsara:136402}.  Since we will see that the Mott transition occurs at {\sl smaller} $U/t$ with increasing $\lambda/t$, we expect that the slave-rotor approximation should be reasonable to describe it for the full range of SOI.  It clearly fails at large $U/t$, but we can substitute direct analysis of the spin-orbital model in that limit.

We decompose the physical electron annihilation operator as $d_{Ri\a}=e^{-i\t_{Ri}}f_{Ri\a}$, where the angle $\t_{Ri}$ is the conjugated variable to the number of electrons on site $R,i$ (the ``angular momentum'' of the rotor), and the ``spinon'' $f_{Ri\a}$ carries the rest of the degrees of freedom. The constraint $L_{Ri}=\sum_{\a}f^\dagger_{Ri\a}f^{\vphantom\dagger}_{Ri\a}-n_d$, restricting the physical part of the Hilbert space, is treated on average.  Further, we use the mean-field decomposition of the hopping term that couples the spinons and rotors according to $AB\rightarrow A\av{B}+B\av{A}$. This Mean Field Theory (MFT) reduces the Hamiltonian~(\ref{eq: Hamiltonian}) to two uncoupled Hamiltonians for spinons and rotors:
\begin{eqnarray}
  H_{f}&=&\sum_{Ri\a}(\ve_\a-\mu-h)f^\dagger_{Ri\a}f^{\vphantom\dagger}_{Ri\a}+t\, Q_f\sum_{\substack{\av{Ri,R'i'}\\ \a\a'}} T^{ii'}_{\a\a'}f^\dagger_{Ri\a}f^{\vphantom\dagger}_{R'i'\a'}\nonumber\\
  H_{\theta}&=&\frac{U}{2}\sum_{Ri}L^2_{Ri}+h(L_{Ri}+n_d)
  +t \,Q_\theta\sum_{\av{Ri,R'i'}}e^{i\theta_{Ri}-i\theta_{R'i'}}.
\end{eqnarray}
Here $L_{Ri}=-i\frac{\partial}{\partial\theta_{Ri}}$, the
coordinate-independent Lagrange multiplier $h$ is introduced to treat
the constraint on the angular momentum, and the couplings $Q_f$ and
$Q_\theta$ need to be determined self-consistently from
$Q_f=\av{e^{i\theta_{Ri}-i\theta_{Ri'}}}$, $Q_\theta=\sum_{\a\a'}
T^{ii'}_{\a\a'}\av{f^\dagger_{Ri\a}f^{\vphantom\dagger}_{Ri'\a'}}$ (note
$Q_\theta \leq 0$ in the self-consistent solution). Here we have made
the so-called ``uniform'' mean-field approximation, on the grounds that
it is the one which smoothly connects to the $U/t=0$ limit, and hence
should be appropriate for small to intermediate $U/t$, where, anyway, we
apply the approximation. Note that the strength of the spin-orbit
interaction, $\lambda$, is not renormalized. After the mean-field
decomposition, the spinon and rotor sectors can be solved independently.

We first consider the spinon Hamiltonian, $H_f$.  It is identical to the non-interacting electron Hamiltonian, but with renormalized hopping $t_{\rm eff} = t Q_f$. All the preceding analysis carried out for $U=0$ can therefore be carried over with this replacement.  Since $Q_f<1$, we indeed observe the dimensionless spin-orbit strength $\lambda/(Q_f t)$ is enhanced by correlations.  Physically, however, we must take care that the $f$ fermions are spinons and therefore their properties do not necessarily translate directly to the physical electrons.

We now turn to the charge (rotor) sector.  $H_\theta$ describes rotor bosons moving on the pyrochlore lattice.  The parameter $h$ must be fixed by charge neutrality, $\langle f^\dagger_{Ri\a}f^{\vphantom\dagger}_{Ri\a}\rangle = n_d$, and hence $\langle L_{Ri}\rangle=0$.  We therefore take $h=0$, which guarantees the latter condition, since then $H_\theta$ has particle-hole symmetry, $L_{Ri}\rightarrow -L_{Ri}, \theta_{Ri}\rightarrow -\theta_{Ri}$.  $H_\theta$ is then expected to exhibit two phases.  For $U/( Q_\theta t) \ll 1$, the rotors are condensed, $\langle e^{i\theta_{Ri}}\rangle \neq 0$, while for $U/(t Q_\theta) \gg 1$, they form an uncondensed Bose Mott insulator with a gap and $\langle e^{i\theta_{Ri}}\rangle =0$.  In the rotor condensate, the electron and spinon operators are proportional, and one has ``Fermi liquid'' states where the spinon band structure describes physical electrons.  When the rotor is uncondensed, there is spin-charge separation and the spinons are emergent neutral quasiparticles carrying spin only.  Separating these two states is expected to be a continuous Quantum Critical Point (QCP).

Combining the solutions of the spin and charge sectors, one finds four possible phases.  When the rotors are condensed, we obtain metallic and TBI phases.  When the rotors are uncondensed, the corresponding phases are ``exotic'' non-magnetic insulators.  For small $\lambda/(Q_f t)$, this is a gapless insulator with a spinon Fermi surface.  For large $\lambda/(Q_f t)$ there is a bulk gap, but the spinons have the non-trivial TBI topolgy; we call this the ``Topological Mott Insulator'' (TMI) phase.  Its properties will be discussed further below.

To determine the boundaries between these phases, we must make MFT self-consistent. From the solution of $H_f$, we can obtain one self-consistent condition $Q_\theta = f(\frac{\lambda}{ Q_ft})$, with $f(\xi)$ determined numerically.  The solution of $H_\theta$ in turn fixes the other self-consistent equation, $Q_f = g(\frac{U}{ Q_\theta t})$.  In principle, $g(\xi)$ (which decreases with increasing $\xi$ from $1$ to $0$) could be determined numerically, since $H_\theta$ is amenable to quantum Monte Carlo without a sign problem.  However, we content ourselves with a simple analytical approximation which produces reasonable results. Further details on the selfconsistent solution of MFT to extract the phase boundaries $\lambda(U)$ are given in the Supplementary Information. Here we present the resulting phase diagram only, Fig.~\ref{fig:phased}.

{\em Discussion: } The combined predictions of the slave-rotor theory
and the strong coupling limit are shown in the phase diagram of
Fig.~\ref{fig:phased}.   Focusing first on weak to intermediate correlation, the theory predicts two types of metal-insulator transitions.  The conventional Mott transition in which the insulator is of Mott type -- the Gapless Mott Insulator (GMI) -- occurs only for rather weak SOI.  Instead, the dominant metal-insulator transition is actually to TBI rather than a Mott insulator, due to the bandwidth reduction with increasing $U$.  This is not a Mott transition per se, as the TBI is qualitatively described by band theory.  Instead, the true Mott transition occurs within the bulk insulating state, between the TBI and the Topological Mott Insulator (TMI).  In the TMI, the charge is stripped from the spinons, resulting most notably in a change from a conducting surface to an insulating one, which nevertheless still supports gapless spin excitations.

The above slave-rotor MFT, by construction, leads to exotic Mott
insulators (the GMI, TMI) with spin-charge separation.  More precisely,
these states are {\em U(1) spin liquids}, which, when fluctuations are
properly taken into account, must support an emergent dynamical U(1)
gauge field.  Such spin liquid states are known to be stable in three
dimensions.  In addition to the excitations already described, they
support gapped ``magnetic monopoles'', which are sources of the emergent
magnetic flux.  Such a monopole in a topological insulator carries a
quantized gauge
charge~\cite{qi2009inducing,qi2008topological,essin2009magnetoelectric,wilczek1987two}.
Deep in the TMI, where charge excitations are costly, we thus expect a
spinon to be bound to the monopole, i.e. the monopole has spin.  The low
energy effects of gauge fluctuations in the GMI (and near the metal-GMI
transition) have been discussed elsewhere (see
Ref.~\cite{podolsky-2009-102}).  We expect only moderate effects of gauge
fluctuations on the gapless surface spinons in the TMI state, owing to
the three-dimensional character of the gauge field.

Is the mean-field starting point itself reasonable?  Because in these
phases the spinons behave very similarly to electrons in the nearby
(metallic and TBI) conventional states, they are most natural in
``weak'' Mott insulators, in which there are significant charge
fluctuations.  This is in line with the MFT prediction of continuous
Mott transitions for these phases.  In reality, many effects beyond MFT
might lead to these transitions becoming first-order, as often observed
in experiment.  However, numerous recent  analytical and numerical
studies~\cite{senthil2008theory,kyung:046402,morita2002nonmagnetic,lee2009mott,yoshikawa:144429,mizusaki:014421} support the existence of continuous or only weakly
discontinuous Mott transitions in {\sl frustrated} geometries such as
the one studied here.  Because the Mott
transition is suppressed to small $U/t$ by increasing SOI, we believe
the predictions of the slave-rotor MFT are reasonable both for the GMI
and the TMI phases.

Turning specifically to the character of the weak to intermediate
coupling transitions in the phase diagram, we expect that effects beyond
MFT will indeed make several of them weakly first order.  First, the
metal-TBI transition, which in MFT is described by a simple crossing of
valence and conduction bands at the $\Gamma$ and $L$ points in the Brillouin zone, respectively, is modified by the effects of long-range Coulomb interactions. This leads to exciton
condensation, occupying a region of the phase diagram in the vicinity of
the putative metal-TBI transition~\cite{RevModPhys.40.755}. If
electron-phonon coupling is not too strong, the most likely excitonic
condensation is predominantly triplet, and this state should evince weak
magnetic order~\cite{RevModPhys.40.755}. Dependent upon the details of
the magnetic order, band structure, and disorder, the transitions to
this state might be first or second order.  Second, the TBI-TMI
transition, described in MFT by the condensation of the rotor, is
modified by coupling to the emergent U(1) gauge field in the TMI state.
This problem maps onto the abelian Higgs transition in 3+1=4 dimensions,
where it can be studied by renormalization group methods.  The gauge
fluctuations destabilize the continuous mean-field transition, rendering
it weakly first order~\cite{PhysRevLett.32.292}. The metal-GMI transition
has been studied recently in Ref.~\cite{podolsky-2009-102}, and is
potentially continuous even with fluctuations.  We do not discuss the
GMI-TMI transition here.

The slave rotor approximation neglects possible magnetic instabilities,
which become increasingly prominent with increasing $U/t$.  For
comparison, in Ref.~\cite{JYucondmat}, a phase diagram in the same
parameter space was obtained for the layered material Na$_{2}$IrO$_{3}$,
based on density functional theory, in which the Mott phases are all
magnetic.  None of the phases we have found are perturbatively unstable
to magnetic order (excepting the excitonic region described above), so
the simplest scenario is one in which a direct transition occurs at
larger $U/t$ from the GMI/TMI states to magnetically ordered ones
(e.g. along the dotted line in Fig.~\ref{fig:phased}).  It will be
interesting to consider whether this might occur by condensation of the
monopole excitations.  It is also
possible for magnetism to occur at smaller $U/t$ for other reasons.  For
instance, nesting in the metallic state would promote spin density wave
(SDW) formation, which could occur by a Slater transition below the Mott
transition predicted by the slave rotor approximation.  In our
calculations, we do not see obviously nesting in the metallic Fermi
surfaces, but this could occur if other hopping processes were
important.  Another interesting possibility is emergence of magnetic
ordering at the {\sl surface} of the TBI or TMI states, where the
surface fermions can undergo Slater transitions.

We conclude with a discussion of experiments and materials.  Experiments
on Ir pyrochlores, Ln$_2$Ir$_2$O$_7$ show a zero-temperature metal
insulator transition with increasing Ln radius, occuring between Ln=Nd
and Ln=Pr~\cite{matsuhira2007metal}.  The insulating materials show a
weak continuous $T>0$ metal-insulator transition, with some magnetic
order evident in the low-temperature insulating state.  This might be
consistent with the excitonic state expected near the metal--TBI
boundary.  However, numerous other interpretations are possible, and
clearly more theoretical and experimental work is required to better
understand these materials.  More generally, the form of the phase
diagram in Fig.~\ref{fig:phased} may apply to a variety of 4d and 5d
transition metal compounds in which both spin-orbit coupling and
correlations play important roles, and whose lattice structures are
favorable for TBI physics in the weak correlation limit.  In Ir$^{4+}$
materials, this requires Ir-O-Ir bond angles far from 180$^\circ$ (for
180$^\circ$ bonds crucial complex phases in the $j=1/2$ wavefunctions
cancel).  Our pyrochlore model here is an example, as is the
two-dimensional honeycomb model considered in
Ref.~\cite{shitade:256403}. Even when TBI physics is absent, the strong
enhancement of spin-orbit physics by correlations must play an important
role in a great many materials.


This work was supported by the DOE through Basic Energy Sciences grants
DE-FG02-08ER46524 (L.B.) and DEFG02-07ER46452 (D.A.P.). The research
facilities at the KITP were supported by the National Science Foundation
grant NSF PHY-0551164.

\bibliography{Ir}

\begin{thebibliography}{36}
\expandafter\ifx\csname natexlab\endcsname\relax\def\natexlab#1{#1}\fi
\expandafter\ifx\csname bibnamefont\endcsname\relax
  \def\bibnamefont#1{#1}\fi
\expandafter\ifx\csname bibfnamefont\endcsname\relax
  \def\bibfnamefont#1{#1}\fi
\expandafter\ifx\csname citenamefont\endcsname\relax
  \def\citenamefont#1{#1}\fi
\expandafter\ifx\csname url\endcsname\relax
  \def\url#1{\texttt{#1}}\fi
\expandafter\ifx\csname urlprefix\endcsname\relax\def\urlprefix{URL }\fi
\providecommand{\bibinfo}[2]{#2}
\providecommand{\eprint}[2][]{\url{#2}}

\bibitem[{\citenamefont{Kane and Mele}(2005)}]{KaneMeleA}
\bibinfo{author}{\bibfnamefont{C.}~\bibnamefont{Kane}} \bibnamefont{and}
  \bibinfo{author}{\bibfnamefont{E.}~\bibnamefont{Mele}},
  \bibinfo{journal}{Phys. Rev. Lett.} \textbf{\bibinfo{volume}{95}},
  \bibinfo{pages}{146802} (\bibinfo{year}{2005}).

\bibitem[{\citenamefont{Bernevig et~al.}(2006)\citenamefont{Bernevig, Hughes,
  and Zhang}}]{citeulike:5096324}
\bibinfo{author}{\bibfnamefont{B.~A.} \bibnamefont{Bernevig}},
  \bibinfo{author}{\bibfnamefont{T.~L.} \bibnamefont{Hughes}},
  \bibnamefont{and} \bibinfo{author}{\bibfnamefont{S.-C.} \bibnamefont{Zhang}},
  \bibinfo{journal}{Science} \textbf{\bibinfo{volume}{314}},
  \bibinfo{pages}{1757} (\bibinfo{year}{2006}).

\bibitem[{\citenamefont{Moore and Balents}({2007})}]{BalentsMoore}
\bibinfo{author}{\bibfnamefont{J.~E.} \bibnamefont{Moore}} \bibnamefont{and}
  \bibinfo{author}{\bibfnamefont{L.}~\bibnamefont{Balents}},
  \bibinfo{journal}{{Phys. Rev. B}} \textbf{\bibinfo{volume}{{75}}},
  \bibinfo{pages}{{121306}} (\bibinfo{year}{{2007}}).

\bibitem[{\citenamefont{Fu et~al.}({2007})\citenamefont{Fu, Kane, and
  Mele}}]{FuKaneMele3D}
\bibinfo{author}{\bibfnamefont{L.}~\bibnamefont{Fu}},
  \bibinfo{author}{\bibfnamefont{C.~L.} \bibnamefont{Kane}}, \bibnamefont{and}
  \bibinfo{author}{\bibfnamefont{E.~J.} \bibnamefont{Mele}},
  \bibinfo{journal}{{Phys. Rev. Lett.}} \textbf{\bibinfo{volume}{{98}}},
  \bibinfo{pages}{{106803}} (\bibinfo{year}{{2007}}).

\bibitem[{\citenamefont{Konig et~al.}(2007)\citenamefont{Konig, Wiedmann,
  Brune, Roth, Buhmann, Molenkamp, Qi, and Zhang}}]{Molenkamp}
\bibinfo{author}{\bibfnamefont{M.}~\bibnamefont{Konig}},
  \bibinfo{author}{\bibfnamefont{S.}~\bibnamefont{Wiedmann}},
  \bibinfo{author}{\bibfnamefont{C.}~\bibnamefont{Brune}},
  \bibinfo{author}{\bibfnamefont{A.}~\bibnamefont{Roth}},
  \bibinfo{author}{\bibfnamefont{H.}~\bibnamefont{Buhmann}},
  \bibinfo{author}{\bibfnamefont{L.~W.} \bibnamefont{Molenkamp}},
  \bibinfo{author}{\bibfnamefont{X.-L.} \bibnamefont{Qi}}, \bibnamefont{and}
  \bibinfo{author}{\bibfnamefont{S.-C.} \bibnamefont{Zhang}},
  \bibinfo{journal}{Science} \textbf{\bibinfo{volume}{318}},
  \bibinfo{pages}{766} (\bibinfo{year}{2007}).

\bibitem[{\citenamefont{Hsieh et~al.}({2008})\citenamefont{Hsieh, Qian, Wray,
  Xia, Hor, Cava, and Hasan}}]{ZahidHasan}
\bibinfo{author}{\bibfnamefont{D.}~\bibnamefont{Hsieh}},
  \bibinfo{author}{\bibfnamefont{D.}~\bibnamefont{Qian}},
  \bibinfo{author}{\bibfnamefont{L.}~\bibnamefont{Wray}},
  \bibinfo{author}{\bibfnamefont{Y.}~\bibnamefont{Xia}},
  \bibinfo{author}{\bibfnamefont{Y.~S.} \bibnamefont{Hor}},
  \bibinfo{author}{\bibfnamefont{R.~J.} \bibnamefont{Cava}}, \bibnamefont{and}
  \bibinfo{author}{\bibfnamefont{M.~Z.} \bibnamefont{Hasan}},
  \bibinfo{journal}{{Nature}} \textbf{\bibinfo{volume}{{452}}},
  \bibinfo{pages}{970} (\bibinfo{year}{{2008}}).

\bibitem[{\citenamefont{Chen and Balents}({2008})}]{GangLB}
\bibinfo{author}{\bibfnamefont{G.}~\bibnamefont{Chen}} \bibnamefont{and}
  \bibinfo{author}{\bibfnamefont{L.}~\bibnamefont{Balents}},
  \bibinfo{journal}{{Phys. Rev. B}} \textbf{\bibinfo{volume}{{78}}},
  \bibinfo{pages}{{094403}} (\bibinfo{year}{{2008}}).

\bibitem[{\citenamefont{Tovar et~al.}(2009)\citenamefont{Tovar, Raman, and
  Shtengel}}]{tovar:024405}
\bibinfo{author}{\bibfnamefont{M.}~\bibnamefont{Tovar}},
  \bibinfo{author}{\bibfnamefont{K.~S.} \bibnamefont{Raman}}, \bibnamefont{and}
  \bibinfo{author}{\bibfnamefont{K.}~\bibnamefont{Shtengel}},
  \bibinfo{journal}{Phys. Rev. B} \textbf{\bibinfo{volume}{79}},
  \bibinfo{eid}{024405} (\bibinfo{year}{2009}).

\bibitem[{\citenamefont{Chen et~al.}({2009})\citenamefont{Chen, Balents, and
  Schnyder}}]{GangSchnyderLB}
\bibinfo{author}{\bibfnamefont{G.}~\bibnamefont{Chen}},
  \bibinfo{author}{\bibfnamefont{L.}~\bibnamefont{Balents}}, \bibnamefont{and}
  \bibinfo{author}{\bibfnamefont{A.~P.} \bibnamefont{Schnyder}},
  \bibinfo{journal}{{Phys. Rev. Lett.}} \textbf{\bibinfo{volume}{{102}}},
  \bibinfo{pages}{{096406}} (\bibinfo{year}{{2009}}).

\bibitem[{\citenamefont{Kim et~al.}(2009)\citenamefont{Kim, Ohsumi, Komesu,
  Sakai, Morita, Takagi, and Arima}}]{citeulike:4746735}
\bibinfo{author}{\bibfnamefont{B.~J.} \bibnamefont{Kim}},
  \bibinfo{author}{\bibfnamefont{H.}~\bibnamefont{Ohsumi}},
  \bibinfo{author}{\bibfnamefont{T.}~\bibnamefont{Komesu}},
  \bibinfo{author}{\bibfnamefont{S.}~\bibnamefont{Sakai}},
  \bibinfo{author}{\bibfnamefont{T.}~\bibnamefont{Morita}},
  \bibinfo{author}{\bibfnamefont{H.}~\bibnamefont{Takagi}}, \bibnamefont{and}
  \bibinfo{author}{\bibfnamefont{T.}~\bibnamefont{Arima}},
  \bibinfo{journal}{Science} \textbf{\bibinfo{volume}{323}},
  \bibinfo{pages}{1329} (\bibinfo{year}{2009}).

\bibitem[{\citenamefont{Shitade et~al.}(2009)\citenamefont{Shitade, Katsura,
  Kune\v{s}, Qi, Zhang, and Nagaosa}}]{shitade:256403}
\bibinfo{author}{\bibfnamefont{A.}~\bibnamefont{Shitade}},
  \bibinfo{author}{\bibfnamefont{H.}~\bibnamefont{Katsura}},
  \bibinfo{author}{\bibfnamefont{J.}~\bibnamefont{Kune\v{s}}},
  \bibinfo{author}{\bibfnamefont{X.-L.} \bibnamefont{Qi}},
  \bibinfo{author}{\bibfnamefont{S.-C.} \bibnamefont{Zhang}}, \bibnamefont{and}
  \bibinfo{author}{\bibfnamefont{N.}~\bibnamefont{Nagaosa}},
  \bibinfo{journal}{Phys. Rev. Lett.} \textbf{\bibinfo{volume}{102}},
  \bibinfo{eid}{256403} (\bibinfo{year}{2009}).

\bibitem[{\citenamefont{Matsuhira et~al.}(2007)\citenamefont{Matsuhira,
  Wakeshima, Nakanishi, Yamada, Nakamura, Kawano, Takagi, and
  Hinatsu}}]{matsuhira2007metal}
\bibinfo{author}{\bibfnamefont{K.}~\bibnamefont{Matsuhira}},
  \bibinfo{author}{\bibfnamefont{M.}~\bibnamefont{Wakeshima}},
  \bibinfo{author}{\bibfnamefont{R.}~\bibnamefont{Nakanishi}},
  \bibinfo{author}{\bibfnamefont{T.}~\bibnamefont{Yamada}},
  \bibinfo{author}{\bibfnamefont{A.}~\bibnamefont{Nakamura}},
  \bibinfo{author}{\bibfnamefont{W.}~\bibnamefont{Kawano}},
  \bibinfo{author}{\bibfnamefont{S.}~\bibnamefont{Takagi}}, \bibnamefont{and}
  \bibinfo{author}{\bibfnamefont{Y.}~\bibnamefont{Hinatsu}},
  \bibinfo{journal}{J. Phys. Soc. Jpn.} \textbf{\bibinfo{volume}{76}},
  \bibinfo{pages}{043706} (\bibinfo{year}{2007}).

\bibitem[{\citenamefont{Okamoto et~al.}(2007)\citenamefont{Okamoto, Nohara,
  Aruga-Katori, and Takagi}}]{okamoto2007spin}
\bibinfo{author}{\bibfnamefont{Y.}~\bibnamefont{Okamoto}},
  \bibinfo{author}{\bibfnamefont{M.}~\bibnamefont{Nohara}},
  \bibinfo{author}{\bibfnamefont{H.}~\bibnamefont{Aruga-Katori}},
  \bibnamefont{and} \bibinfo{author}{\bibfnamefont{H.}~\bibnamefont{Takagi}},
  \bibinfo{journal}{Phys. Rev. Lett.} \textbf{\bibinfo{volume}{99}},
  \bibinfo{pages}{137207} (\bibinfo{year}{2007}).

\bibitem[{\citenamefont{Fukazawa and Maeno}(2002)}]{fukazawa2002filling}
\bibinfo{author}{\bibfnamefont{H.}~\bibnamefont{Fukazawa}} \bibnamefont{and}
  \bibinfo{author}{\bibfnamefont{Y.}~\bibnamefont{Maeno}}, \bibinfo{journal}{J.
  Phys. Soc. Jpn.} \textbf{\bibinfo{volume}{71}}, \bibinfo{pages}{2578}
  (\bibinfo{year}{2002}).

\bibitem[{\citenamefont{Nakatsuji et~al.}(2006)\citenamefont{Nakatsuji,
  Machida, Maeno, Tayama, Sakakibara, Duijn, Balicas, Millican, Macaluso, and
  Chan}}]{nakatsuji2006metallic}
\bibinfo{author}{\bibfnamefont{S.}~\bibnamefont{Nakatsuji}},
  \bibinfo{author}{\bibfnamefont{Y.}~\bibnamefont{Machida}},
  \bibinfo{author}{\bibfnamefont{Y.}~\bibnamefont{Maeno}},
  \bibinfo{author}{\bibfnamefont{T.}~\bibnamefont{Tayama}},
  \bibinfo{author}{\bibfnamefont{T.}~\bibnamefont{Sakakibara}},
  \bibinfo{author}{\bibfnamefont{J.}~\bibnamefont{Duijn}},
  \bibinfo{author}{\bibfnamefont{L.}~\bibnamefont{Balicas}},
  \bibinfo{author}{\bibfnamefont{J.}~\bibnamefont{Millican}},
  \bibinfo{author}{\bibfnamefont{R.}~\bibnamefont{Macaluso}}, \bibnamefont{and}
  \bibinfo{author}{\bibfnamefont{J.}~\bibnamefont{Chan}},
  \bibinfo{journal}{Phys. Rev. Lett.} \textbf{\bibinfo{volume}{96}},
  \bibinfo{pages}{87204} (\bibinfo{year}{2006}).

\bibitem[{\citenamefont{Abragam and Bleaney}(1975)}]{Abragam}
\bibinfo{author}{\bibfnamefont{A.}~\bibnamefont{Abragam}} \bibnamefont{and}
  \bibinfo{author}{\bibfnamefont{B.}~\bibnamefont{Bleaney}},
  \emph{\bibinfo{title}{Electron paramagnetic resonance of transition ions}}
  (\bibinfo{publisher}{Oxford, Clarendon Press}, \bibinfo{year}{1975}).

\bibitem[{\citenamefont{Jin et~al.}(2009)\citenamefont{Jin, Kim, Jeong, Kim,
  and Yu}}]{JYucondmat}
\bibinfo{author}{\bibfnamefont{H.}~\bibnamefont{Jin}},
  \bibinfo{author}{\bibfnamefont{H.}~\bibnamefont{Kim}},
  \bibinfo{author}{\bibfnamefont{H.}~\bibnamefont{Jeong}},
  \bibinfo{author}{\bibfnamefont{C.~H.} \bibnamefont{Kim}}, \bibnamefont{and}
  \bibinfo{author}{\bibfnamefont{J.}~\bibnamefont{Yu}},
  \bibinfo{journal}{arXiv:0907.0743}  (\bibinfo{year}{2009}).

\bibitem[{\citenamefont{Slater and Koster}(1954)}]{SlaterKoster}
\bibinfo{author}{\bibfnamefont{J.}~\bibnamefont{Slater}} \bibnamefont{and}
  \bibinfo{author}{\bibfnamefont{G.}~\bibnamefont{Koster}},
  \bibinfo{journal}{Phys. Rev.} \textbf{\bibinfo{volume}{94}},
  \bibinfo{pages}{1498 } (\bibinfo{year}{1954}).

\bibitem[{\citenamefont{Fu and Kane}(2007)}]{FuKaneTBIinv2007}
\bibinfo{author}{\bibfnamefont{L.}~\bibnamefont{Fu}} \bibnamefont{and}
  \bibinfo{author}{\bibfnamefont{C.}~\bibnamefont{Kane}},
  \bibinfo{journal}{Phys. Rev. B} \textbf{\bibinfo{volume}{76}},
  \bibinfo{pages}{045302} (\bibinfo{year}{2007}).

\bibitem[{\citenamefont{Elhajal et~al.}(2005)\citenamefont{Elhajal, Canals,
  Sunyer, and Lacroix}}]{Elhajal}
\bibinfo{author}{\bibfnamefont{M.}~\bibnamefont{Elhajal}},
  \bibinfo{author}{\bibfnamefont{B.}~\bibnamefont{Canals}},
  \bibinfo{author}{\bibfnamefont{R.}~\bibnamefont{Sunyer}}, \bibnamefont{and}
  \bibinfo{author}{\bibfnamefont{C.}~\bibnamefont{Lacroix}},
  \bibinfo{journal}{Phys. Rev. B} \textbf{\bibinfo{volume}{71}},
  \bibinfo{pages}{94420} (\bibinfo{year}{2005}).

\bibitem[{\citenamefont{Florens and Georges}({2004})}]{FlorensGeorges}
\bibinfo{author}{\bibfnamefont{S.}~\bibnamefont{Florens}} \bibnamefont{and}
  \bibinfo{author}{\bibfnamefont{A.}~\bibnamefont{Georges}},
  \bibinfo{journal}{{Phys. Rev. B}} \textbf{\bibinfo{volume}{{70}}},
  \bibinfo{pages}{{035114}} (\bibinfo{year}{{2004}}).

\bibitem[{\citenamefont{Mizusaki and Imada}(2006)}]{mizusaki:014421}
\bibinfo{author}{\bibfnamefont{T.}~\bibnamefont{Mizusaki}} \bibnamefont{and}
  \bibinfo{author}{\bibfnamefont{M.}~\bibnamefont{Imada}},
  \bibinfo{journal}{Phys. Rev. B} \textbf{\bibinfo{volume}{74}},
  \bibinfo{eid}{014421} (\bibinfo{year}{2006}).

\bibitem[{\citenamefont{Motrunich}(2005)}]{motrunich2005vst}
\bibinfo{author}{\bibfnamefont{O.}~\bibnamefont{Motrunich}},
  \bibinfo{journal}{Phys. Rev. B} \textbf{\bibinfo{volume}{72}},
  \bibinfo{pages}{045105} (\bibinfo{year}{2005}).

\bibitem[{\citenamefont{Sahebsara and
  S\'{e}n\'{e}chal}(2008)}]{sahebsara:136402}
\bibinfo{author}{\bibfnamefont{P.}~\bibnamefont{Sahebsara}} \bibnamefont{and}
  \bibinfo{author}{\bibfnamefont{D.}~\bibnamefont{S\'{e}n\'{e}chal}},
  \bibinfo{journal}{Phys. Rev. Lett.} \textbf{\bibinfo{volume}{100}},
  \bibinfo{eid}{136402} (\bibinfo{year}{2008}).

\bibitem[{\citenamefont{Qi et~al.}(2009)\citenamefont{Qi, Li, Zang, and
  Zhang}}]{qi2009inducing}
\bibinfo{author}{\bibfnamefont{X.}~\bibnamefont{Qi}},
  \bibinfo{author}{\bibfnamefont{R.}~\bibnamefont{Li}},
  \bibinfo{author}{\bibfnamefont{J.}~\bibnamefont{Zang}}, \bibnamefont{and}
  \bibinfo{author}{\bibfnamefont{S.}~\bibnamefont{Zhang}},
  \bibinfo{journal}{Science} \textbf{\bibinfo{volume}{323}},
  \bibinfo{pages}{1184} (\bibinfo{year}{2009}).

\bibitem[{\citenamefont{Qi et~al.}(2008)\citenamefont{Qi, Hughes, and
  Zhang}}]{qi2008topological}
\bibinfo{author}{\bibfnamefont{X.}~\bibnamefont{Qi}},
  \bibinfo{author}{\bibfnamefont{T.}~\bibnamefont{Hughes}}, \bibnamefont{and}
  \bibinfo{author}{\bibfnamefont{S.}~\bibnamefont{Zhang}},
  \bibinfo{journal}{Phys. Rev. B} \textbf{\bibinfo{volume}{78}},
  \bibinfo{pages}{195424} (\bibinfo{year}{2008}).

\bibitem[{\citenamefont{Essin et~al.}(2009)\citenamefont{Essin, Moore, and
  Vanderbilt}}]{essin2009magnetoelectric}
\bibinfo{author}{\bibfnamefont{A.}~\bibnamefont{Essin}},
  \bibinfo{author}{\bibfnamefont{J.}~\bibnamefont{Moore}}, \bibnamefont{and}
  \bibinfo{author}{\bibfnamefont{D.}~\bibnamefont{Vanderbilt}},
  \bibinfo{journal}{Phys. Rev. Lett.} \textbf{\bibinfo{volume}{102}},
  \bibinfo{pages}{146805} (\bibinfo{year}{2009}).

\bibitem[{\citenamefont{Wilczek}(1987)}]{wilczek1987two}
\bibinfo{author}{\bibfnamefont{F.}~\bibnamefont{Wilczek}},
  \bibinfo{journal}{Phys. Rev. Lett.} \textbf{\bibinfo{volume}{58}},
  \bibinfo{pages}{1799} (\bibinfo{year}{1987}).

\bibitem[{\citenamefont{Podolsky et~al.}(2009)\citenamefont{Podolsky,
  Paramekanti, Kim, and Senthil}}]{podolsky-2009-102}
\bibinfo{author}{\bibfnamefont{D.}~\bibnamefont{Podolsky}},
  \bibinfo{author}{\bibfnamefont{A.}~\bibnamefont{Paramekanti}},
  \bibinfo{author}{\bibfnamefont{Y.~B.} \bibnamefont{Kim}}, \bibnamefont{and}
  \bibinfo{author}{\bibfnamefont{T.}~\bibnamefont{Senthil}},
  \bibinfo{journal}{Phys. Rev. Lett.} \textbf{\bibinfo{volume}{102}},
  \bibinfo{pages}{186401} (\bibinfo{year}{2009}).

\bibitem[{\citenamefont{Senthil}(2008)}]{senthil2008theory}
\bibinfo{author}{\bibfnamefont{T.}~\bibnamefont{Senthil}},
  \bibinfo{journal}{Phys. Rev. B} \textbf{\bibinfo{volume}{78}},
  \bibinfo{pages}{45109} (\bibinfo{year}{2008}).

\bibitem[{\citenamefont{Kyung and Tremblay}(2006)}]{kyung:046402}
\bibinfo{author}{\bibfnamefont{B.}~\bibnamefont{Kyung}} \bibnamefont{and}
  \bibinfo{author}{\bibfnamefont{A.-M.~S.} \bibnamefont{Tremblay}},
  \bibinfo{journal}{Phys. Rev. Lett.} \textbf{\bibinfo{volume}{97}},
  \bibinfo{eid}{046402} (\bibinfo{year}{2006}).

\bibitem[{\citenamefont{Morita et~al.}(2002)\citenamefont{Morita, Watanabe, and
  Imada}}]{morita2002nonmagnetic}
\bibinfo{author}{\bibfnamefont{H.}~\bibnamefont{Morita}},
  \bibinfo{author}{\bibfnamefont{S.}~\bibnamefont{Watanabe}}, \bibnamefont{and}
  \bibinfo{author}{\bibfnamefont{M.}~\bibnamefont{Imada}}, \bibinfo{journal}{J.
  Phys. Soc. Jpn.} \textbf{\bibinfo{volume}{71}}, \bibinfo{pages}{2109}
  (\bibinfo{year}{2002}).

\bibitem[{\citenamefont{Lee and Monien}(2009)}]{lee2009mott}
\bibinfo{author}{\bibfnamefont{H.}~\bibnamefont{Lee}} \bibnamefont{and}
  \bibinfo{author}{\bibfnamefont{H.}~\bibnamefont{Monien}},
  \bibinfo{journal}{arXiv:0903.3005}  (\bibinfo{year}{2009}).

\bibitem[{\citenamefont{Yoshikawa and Ogata}(2009)}]{yoshikawa:144429}
\bibinfo{author}{\bibfnamefont{T.}~\bibnamefont{Yoshikawa}} \bibnamefont{and}
  \bibinfo{author}{\bibfnamefont{M.}~\bibnamefont{Ogata}},
  \bibinfo{journal}{Phys. Rev. B} \textbf{\bibinfo{volume}{79}},
  \bibinfo{eid}{144429} (pages~\bibinfo{numpages}{7}) (\bibinfo{year}{2009}).

\bibitem[{\citenamefont{Halperin and Rice}(1968)}]{RevModPhys.40.755}
\bibinfo{author}{\bibfnamefont{B.~I.} \bibnamefont{Halperin}} \bibnamefont{and}
  \bibinfo{author}{\bibfnamefont{T.~M.} \bibnamefont{Rice}},
  \bibinfo{journal}{Rev. Mod. Phys.} \textbf{\bibinfo{volume}{40}},
  \bibinfo{pages}{755} (\bibinfo{year}{1968}).

\bibitem[{\citenamefont{Halperin et~al.}(1974)\citenamefont{Halperin, Lubensky,
  and Ma}}]{PhysRevLett.32.292}
\bibinfo{author}{\bibfnamefont{B.~I.} \bibnamefont{Halperin}},
  \bibinfo{author}{\bibfnamefont{T.~C.} \bibnamefont{Lubensky}},
  \bibnamefont{and} \bibinfo{author}{\bibfnamefont{S.-k.} \bibnamefont{Ma}},
  \bibinfo{journal}{Phys. Rev. Lett.} \textbf{\bibinfo{volume}{32}},
  \bibinfo{pages}{292} (\bibinfo{year}{1974}).

\end{thebibliography}


\begin{thebibliography}{2}
\expandafter\ifx\csname natexlab\endcsname\relax\def\natexlab#1{#1}\fi
\expandafter\ifx\csname bibnamefont\endcsname\relax
  \def\bibnamefont#1{#1}\fi
\expandafter\ifx\csname bibfnamefont\endcsname\relax
  \def\bibfnamefont#1{#1}\fi
\expandafter\ifx\csname citenamefont\endcsname\relax
  \def\citenamefont#1{#1}\fi
\expandafter\ifx\csname url\endcsname\relax
  \def\url#1{\texttt{#1}}\fi
\expandafter\ifx\csname urlprefix\endcsname\relax\def\urlprefix{URL }\fi
\providecommand{\bibinfo}[2]{#2}
\providecommand{\eprint}[2][]{\url{#2}}

\bibitem[{\citenamefont{Fu and Kane}(2007)}]{FuKaneTBIinv2007}
\bibinfo{author}{\bibfnamefont{L.}~\bibnamefont{Fu}} \bibnamefont{and}
  \bibinfo{author}{\bibfnamefont{C.}~\bibnamefont{Kane}},
  \bibinfo{journal}{Phys. Rev. B} \textbf{\bibinfo{volume}{76}},
  \bibinfo{pages}{045302} (\bibinfo{year}{2007}).

\bibitem[{\citenamefont{Florens and Georges}({2004})}]{FlorensGeorges}
\bibinfo{author}{\bibfnamefont{S.}~\bibnamefont{Florens}} \bibnamefont{and}
  \bibinfo{author}{\bibfnamefont{A.}~\bibnamefont{Georges}},
  \bibinfo{journal}{{Phys. Rev. B}} \textbf{\bibinfo{volume}{{70}}},
  \bibinfo{pages}{{035114}} (\bibinfo{year}{{2004}}).

\end{thebibliography}

\end{document}


\title{Supplementary Information for ``Mott physics and band topology in materials with strong spin-orbit interaction''}

\date{\today}

\author{D. A. Pesin}
\email{pesin@physics.utexas.edu}
\altaffiliation[\newline Present address~: ]{Department of Physics, The University of Texas at Austin, Austin, TX 78712, USA}
\affiliation{Department of Physics, University of Washington,
Seattle, WA 98195, USA}
\affiliation{Kavli Institute for Theoretical Physics, Santa Barbara, CA 93106}
\author{Leon Balents}
\affiliation{Kavli Institute for Theoretical Physics, Santa Barbara, CA 93106}
\maketitle

\section{Spin-orbit eigenstates}
The explicit expressions for the wave functions of spin-orbit eigenstates ($j=1/2$ doublet and $j=3/2$ quadruplet), are given in Eq~(\ref{eq:SOeigenstates}). The spin projections are taken in the axes bound to oxygen octahedra of each Ir site (called local octahedral axes from now on).
\begin{equation}\label{eq:SOeigenstates}
    \left(\begin{array}{l}
        |\frac{1}{2},\frac{1}{2}\ket\\
        |\frac{1}{2},-\frac{1}{2}\ket\\
        |\frac{3}{2},\frac{3}{2}\ket\\
        |\frac{3}{2},\frac{1}{2}\ket\\
        |\frac{3}{2},-\frac{1}{2}\ket\\
        |\frac{3}{2},-\frac{3}{2}\ket
  \end{array}\right)=
  \left(\begin{array}{cccccc}
    0&\frac{1}{\sqrt{3}}&0&\frac{i}{\sqrt{3}}&\frac{1}{\sqrt{3}}&0\\
    \frac{1}{\sqrt{3}}&0&-\frac{i}{\sqrt{3}}&0&0&-\frac{1}{\sqrt{3}}\\
    -\frac{1}{\sqrt{2}}&0&-\frac{i}{\sqrt{2}}&0&0&0\\
    0&-\frac{1}{\sqrt{6}}&0&-\frac{i}{\sqrt{6}}&\frac{2}{\sqrt{6}}&0\\
    \frac{1}{\sqrt{6}}&0&-\frac{i}{\sqrt{6}}&0&0&\frac{2}{\sqrt{6}}\\
    0&\frac{1}{\sqrt{2}}&0&-\frac{i}{\sqrt{2}}&0&0
  \end{array}\right)
  \left(\begin{array}{l}
    |yz\up\ket\\
    |yz\down\ket\\
    |zx\up\ket\\
    |zx\down\ket\\
    |xy\up\ket\\
    |xy\down\ket
  \end{array}\right)
\end{equation}

\section{Dimensionless hopping matrices}
The hopping matrices are constructed as follows.

Let us first obtain the hopping matrices for hops between the $t_{2g}$ states in local octahedral axes. A pair of Ir sites is connected via an oxygen site nearest to the two Ir ones. As far as the Ir-O coupling is concerned, the possible hopping matrix elements are restricted by symmetry: if O site lies, say, on the z-axis of the octahedral system of coordinates of a particular Ir site, then electrons can only hop from O's $p_y$ onto Ir's $d_{yz}$ orbital, or from O's $p_x$ onto Ir's $d_{zx}$ one. Analogous rules apply to other placements of O sites, and we encode them by introducing the corresponding Ir-O hopping matrices, $\tau^{\pm}_{x,y,z}$, where the subscript says which axis a particular oxygen resides on, and the superscript shows whether it has positive or negative coordinate: Since p-orbitals have odd parity, this changes the sign of the hopping matrix element. Explicitly, these matrices, connecting $p_x,p_y,p_z$ to $d_{yz}, d_{zx}$ and $d_{xy}$ orbitals, read
\begin{equation}
\tau_x^\pm=\pm V_{pd\pi}\left(
\begin{array}{ccc}
 0 & 0 & 0 \\
 0 & 0 & 1 \\
 0 & 1 & 0
\end{array}
\right);\,\,\,
\tau_y^\pm=\pm V_{pd\pi}\left(
\begin{array}{ccc}
 0 & 0 & 1 \\
 0 & 0 & 0 \\
 1 & 0 & 0
\end{array}
\right);\,\,\,
\tau _z^\pm=\pm V_{pd\pi}\left(
\begin{array}{ccc}
 0 & 1 & 0 \\
 1 & 0 & 0 \\
 0 & 0 & 0
\end{array}
\right).
\end{equation}
Here $V_{pd\pi}$ is the common matrix element, parameterizing the strength of the hopping.

To ``integrate out'' the oxygens we need to view their p orbitals from the same system of coordinates. We choose it to be the global cubic coordinate, and introduce matrices $R^{(i)}$ that rotate the cubic axes into the octahedral ones of site $i$:
\begin{equation}
R^{(0)}=\left(
\begin{array}{ccc}
 \frac{2}{3} & -\frac{1}{3} & -\frac{2}{3} \\
 -\frac{1}{3} & \frac{2}{3} & -\frac{2}{3} \\
 \frac{2}{3} & \frac{2}{3} & \frac{1}{3}
\end{array}
\right),\,\,
R^{(1)}=\left(
\begin{array}{ccc}
 \frac{2}{3} & \frac{2}{3} & \frac{1}{3} \\
 -\frac{2}{3} & \frac{1}{3} & \frac{2}{3} \\
 \frac{1}{3} & -\frac{2}{3} & \frac{2}{3}
\end{array}
\right),\,\,
R^{(2)}=\left(
\begin{array}{ccc}
 \frac{1}{3} & -\frac{2}{3} & \frac{2}{3} \\
 \frac{2}{3} & \frac{2}{3} & \frac{1}{3} \\
 -\frac{2}{3} & \frac{1}{3} & \frac{2}{3}
\end{array}
\right),\,\,
R^{(3)}=\left(
\begin{array}{ccc}
 \frac{1}{3} & -\frac{2}{3} & \frac{2}{3} \\
 -\frac{2}{3} & -\frac{2}{3} & -\frac{1}{3} \\
 \frac{2}{3} & -\frac{1}{3} & -\frac{2}{3}
\end{array}
\right).
\end{equation}

The transformation law for p orbitals creation operators is
$p^+_{in}=R^i_{m n}p^+_{m}$, where the index $i=0\ldots 3$ labels sets
of the octahedral coordinates, operators with a single index correspond to p orbital in cubic axes, and $n,m=1,2,3$ correspond to $x,y,z$, respectively. Integrating out the oxygen orbitals, we obtain the hopping matrices between local $t_{2g}$ orbitals:
\begin{equation}
  t^{ii'}=t\,\tau_{ii'}(R^i)^\textrm{T}R^{i'}\tau_{i'i},
\end{equation}
where $\tau_{ii'}$ is the appropriate choice of $\tau$-matrices for hopping from site $i$ onto site $i'$. For example, for hopping between sites 0 and 3 we have $\tau_{03}=\tau^-_z$, and $\tau_{30}=\tau^+_y$.  As mentioned in the main text, $t \propto V^2_{pd\pi}/\D$, where $\D$ is the energy difference between the Ir d and O p states.

Ir-O hopping conserves the electron spin, thus matrices $t^{ii'}$ are identity in the spin space. However, to project on the local spin-orbit eigenstates we would like to view the spins in local coordinate system. To this end we introduce matrices $D^{(i)}$, which are spinor representations of $O(3)$ rotations $R^{(i)}$:
\begin{equation}
  D^{(0)}=\left(
\begin{array}{cc}
 \sqrt{\frac{2}{3}} & \frac{1-i}{\sqrt{6}} \\
 -\frac{1+i}{\sqrt{6}} & \sqrt{\frac{2}{3}}
\end{array}
\right),\,\,
D^{(1)}=\left(
\begin{array}{cc}
 -\frac{2+i}{\sqrt{6}} & -\frac{i}{\sqrt{6}} \\
 -\frac{i}{\sqrt{6}} & -\frac{2-i}{\sqrt{6}}
\end{array}
\right),\,\,
D^{(2)}=\left(
\begin{array}{cc}
 \frac{2-i}{\sqrt{6}} & -\frac{1}{\sqrt{6}} \\
 \frac{1}{\sqrt{6}} & \frac{2+i}{\sqrt{6}}
\end{array}
\right),\,\,
D^{(3)}=\left(
\begin{array}{cc}
 -\frac{i}{\sqrt{6}} & \frac{1-2 i}{\sqrt{6}} \\
 -\frac{1+2 i}{\sqrt{6}} & \frac{i}{\sqrt{6}}
\end{array}
\right).
\end{equation}

Using these matrices, the spin-dependent dimensionless hopping amplitudes describing hopping of electrons with spin projections measured in local axes (index $\sigma$) between local $t_{2g}$ orbitals (index $l$) are written as
\begin{equation}
  \tilde{T}^{ii'}_{l\sigma,l'\sigma'}=\left[\tau_{ii'}(R^i)^\textrm{T}R^{i'}\tau_{i'i}\right]_{ll'}
  \left[\left(D^{(i)}\right)^+D^{(i')}\right]_{\sigma\sigma'}.
\end{equation}

The final step consists in changing the basis to that of the spin-orbit eigenstates, listed in Eq.~(\ref{eq:SOeigenstates}). We denote these basis states as $|\alpha\rangle=\sum_{l\sigma}M_{\alpha,l\sigma}$, where index $\alpha=1\ldots 6$ numbers states from top to bottom in the column in the left hand side of Eq.~(\ref{eq:SOeigenstates}), and each set of the coefficients $M_{\alpha,l\sigma}$ is essentially the $\alpha th$ row of the matrix in the right hand side of Eq.~(\ref{eq:SOeigenstates}). Using these notations, we finally arrive at the sought dimensionless hopping matrices between the spin-orbit eigenstates in local axes:
\begin{equation}
  T^{ii'}_{\alpha,\alpha'}=\sum_{ll',\sigma\sigma'}\left(M^\alpha_{l\sigma}\right)^*M^{\alpha'}_{l'\sigma'}
  \left[\tau_{ii'}(R^i)^\textrm{T}R^{i'}\tau_{i'i}\right]_{ll'}
  \left[\left(D^{(i)}\right)^+D^{(i')}\right]_{\sigma\sigma'}.
\end{equation}
We will not give the explicit matrices here for space reasons.

\section{Fu and Kane criterion for Topological Insulators}

Here we outline the proof that for large $\lambda/t$ the ground state of the system is a strong topological insulator. The analysis proceeds in the following way. According to Fu and Kane~\cite{FuKaneTBIinv2007}, to extract the topological properties of an insulator with inversion symmetry one needs to study the parity eigenvalues of filled bands at time-reversal-invariant momenta (TRIM). Below we use the notations of Ref.~\cite{FuKaneTBIinv2007}. Then the eight TRIM are written as $\Gamma_{l=n_1n_2n_3}=(n_1G_1+n_2G_2+n_3G_3)/2$, where $n_{1,2,3}=0,1$, and $G_{1,2,3}$ are the reciprocal lattice vectors (i.e. those of the BCC lattice, as the direct one is the FCC in our case).

If the parities, $\xi_m(\Gamma_l)$, of the occupied bands (index $m$) are known, then one finds the $Z_2$ topological invariant $\nu$ that distinguishes the strong topological phase from the trivial and weak topological ones as
\begin{equation}\label{eq:nu}
  (-1)^\nu=\prod_l\prod_{m=1}^{2}\xi_{2m}(\Gamma_l).
\end{equation}
Here we took into account that for $\lambda/t\to \infty$ there are eight relevant bands, four of which are filled, so the index $m$ takes on values $1,2$.

If we choose site $0$ as the inversion center (see Fig.~2 of the main text for details), parity acts on a spin-orbit eigenstate centered at basis site $i$ of Bravais lattice site $R$ as $\hat{P}\psi_i(R)=\psi_i(-R-2b_i)$, where $b_i$ are the basis vectors: $b_0=a(000), b_1=a(011)/4, b_2=a(101)/4, b_3=a(110)/4 $, $a$ is the FCC cubic cell lattice parameter. In $k$-space at TRIM the operator of inversion is a diagonal matrix: $P(\Gamma_l)=\textrm{diag}(e^{2i\Gamma_lb_0},e^{2i\Gamma_lb_0},e^{2i\Gamma_lb_1},e^{2i\Gamma_lb_1},
e^{2i\Gamma_lb_2},e^{2i\Gamma_lb_2},e^{2i\Gamma_lb_3},e^{2i\Gamma_lb_3})$. Using this fact and diagonalizing the noninteracting part of the hopping Hamiltonian (see Eq.(1) in the main text), we can easily find the eigenvalues of $\hat{P}$ in the subspace of wave functions corresponding to the occupied bands. The distribution of $\pm 1$ among $\xi_m(\Gamma_i)$ depends on the choice of the inversion center, however their product determining $\nu$, Eq.~(\ref{eq:nu}), is not, and we obtain $\nu=1$. Therefore, we conclude that we deal with a strong topological insulator for large enough $\lambda$.

The other three (``weak'') $Z_2$ topological invariants, $(\nu_1,\nu_2,\nu_3)$, can also be obtained knowing $\xi_m$'s. While $\nu$ is given by a product of $\xi_{2m}$ evaluated at all TRIM, each of $\nu_i$ is determined by a product over those TRIM that reside in the same plane~\cite{FuKaneTBIinv2007}:
\begin{equation}
  (-1)^{\nu_i}=\prod_{n_i=1,n_{k\neq i}=0,1}\prod_{m=1}^{2}\xi_{2m}(\Gamma_{l=n_1n_2n_3}).
\end{equation}
An explicit calculation gives $\nu_1=\nu_2=\nu_3=0$.

\section{Solution of the selfconsistent MFT}

Following Florens and Georges, we substitute $e^{i\theta_{Ri}}=x_{Ri}+iy_{Ri}$, with the constraint $x_{Ri}^2+y_{Ri}^2=1$, again treated on average by introducing the corresponding Lagrange multiplier, $\rho$.  As explained in the main text, we set the Lagrange multiplier $h=0$, which makes $x_{Ri}$ and $y_{Ri}$ independent. The rotor Hamiltonian, $H_\theta$, reduces to a set of coupled Harmonic oscillators on the pyrochlore lattice. It is easily diagonalized: each of $x_{Ri}$ and $y_{Ri}$ leads to four bands of ``phonons'' that have dispersions $W_{j}(k)=\sqrt{U[\rho+t|Q_\theta|w_{j}(k)]}$, where $w_{1,2}(k)=-2\left(1\pm\sqrt{1+\sum_{\mu<\nu}\cos (k_\mu/2)\cos (k_\nu/2)}\right)$, and $w_{3,4}(k)=2$, such that the former two are dispersive, and the latter are dispersionless. $w_1(k)$ is the lowest band with $\min w_{1}(k)=-6$. Note also that $U$ had to be rescaled, $U\to U/2$, in order to keep the correct atomic limit~\cite{FlorensGeorges}. Having solved the spinon and rotor sectors, we can determine the phase boundaries in a self-consistent manner.

The Mott transition is characterized by the change from gapped, $\rho>6t|Q_\theta|$, to gapless (and Bose-condensed) rotors, $\rho=6t|Q_\theta|$. In the uncondensed phase the equations for $\rho$ and $Q_f$ read, respectively
\begin{align}\label{eq:rho}
&1=\frac{1}{4N_{sites}}\sum_{jk}\sqrt{\frac{U}{4(\rho+t |Q_\theta| w_{j}(k))}},\nonumber\\
&Q_f=-\frac{1}{24N_{sites}}\sum_{jk}w_{jk}\sqrt{\frac{U}{4(\rho+t|Q_\theta|w_{jk})}},
\end{align}
where $N_{sites}$ is the number of sites in the Bravais lattice. These two equations imply that along the line $U_c(\lambda)$, separating the gapped and condensed rotors, $Q_f$ stays constant and equal to
\begin{equation}\label{eq:Qf} Q^c_f=-\frac{\frac{1}{24}\sum_{jk}w_{jk}\frac{1}{\sqrt{6+w_{jk}}}}{\frac{1}{4}\sum_{jk}\frac{1}{\sqrt{6+w_{jk}}}}
  \approx 0.16.
\end{equation}
Having obtained the value of the renormalization factor for the spinon hopping, we then solve the spinon sector to find the chemical potential $\mu(\lambda)$, calculate $Q^c_\theta(\lambda)$, and finally obtain from the first of Eqs.~(\ref{eq:rho}) evaluated at $\rho=6|Q^c_t|$ that  $U_c(Q^c_\theta(\lambda))=4Q^c_\theta(\lambda)\left(\frac{1}{4}\sum_{jk}\frac{1}{\sqrt{6+w_{jk}}}\right)^{-2}\approx 19.5Q^c_\theta(\lambda)$.

To find the line $\lambda_{FS}(U)$ separating  spinons with a Fermi surface from those in the topological insulating state, or its analog on the Mott-insulating part of the phase diagram, we need to generalize equations for $\rho$ and $Q_f$ to account for Bose condensation of rotors in the lowest band, $W_{1}(k)$. Introducing the amplitude of Bose condensation, $Z$, the equations for $\rho$ and $Q_f$ read, respectively:
\begin{eqnarray}
  &&Z+\sqrt{\frac{U}{U_c(Q_\theta)}}=1,\nonumber\\
  &&Q_f=Z+Q^c_f\sqrt{\frac{U}{U_c(Q_\theta)}}.
\end{eqnarray}
To proceed, we first note that on the mean filed level the spinon sector is solved in the same way for condensed and uncondensed rotors, thus we do know that the ratio $\lambda/(Q_ft)$ stays the same along the line separating, say, spinons with a Fermi surface from those in the topological insulating state. And, of course, it is equal to the same ratio for $U=0$ (i.e. it is 2.8). Second, we observe that the value of $Q_\theta$ depends on this very ratio, $\lambda/(Q_ft)$, only. Therefore, $Q_\theta$ stays the same along a line with a given $\lambda/(Q_ft)$. We then solve the spinon problem at $\lambda/(Q_ft)=2.8$ to obtain $|Q_\theta|\approx 0.26$. Then for a given $U$ we find $\rho$ and $Q_f$. The latter determines the sought line simply as $\lambda_{FS}(U)=2.8tQ_f(\frac{U}{Q_\theta t})$. The line corresponding to the gap closure at $\lambda/(Q_ft)\approx 3.3$ is obtained in exactly the same manner.


\bibliography{IrSM}